\documentstyle[prd,aps,epsfig]{revtex}
\tighten
\begin{document}
\draft

\preprint{Local Report No. : UGVA-DPT 1997-05-973}

\twocolumn[\hsize\textwidth\columnwidth\hsize\csname @twocolumnfalse\endcsname

\title{Baryon Density in the Central Region of a Heavy-Ion Collision}
\author{A.J. Gill}
\address{Departement de Physique Theorique, 
Universit\'e de Gen\`eve, Ecole de Physique, \\24, Quai E. Ansermet,
CH-1211 Gen\`eve, Switzerland}
\date{\today}
\maketitle

\begin{abstract}
The O(4) linear sigma model of the chiral transition in QCD is similar to 
models of the superfluid transition in $^4$He and $^3$He. Observations of
vortex formation in superfluid helium have recently improved the understanding
of the dynamics of such transitions. This is exploited to estimate the baryon
density in the central region of a heavy ion region and the result is 
consistent with the long held belief that this density is very small in 
comparison with the pion density.
\end{abstract}


\vskip2pc]

\section{Introduction}

Although QCD is believed to be the underlying theory of the quarks and gluons 
which constitute normal baryonic matter, very little is currently known for
certain about the nature of its chiral phase transition. A purely gluonic
$SU(3)$ theory is believed to have a first order transition \cite{rajagopal}.
In a world with just two massless quarks, however, or in other words an 
infinitely massive strange quark, the chiral transition is plausibly second 
order \cite{wilzcekrajagopal}. If the strange quark were massless the 
transition would again be first order \cite{wilzcekpisarski}. It is even
conceivable that there is no real phase transition at all and that the
change of state occurs by a smooth cross-over \cite{rajagopal}.
For the physical values of the quark masses the nature of the transition is 
not certain although it is frequently assumed to be second order. It is also 
often assumed that the chiral and deconfining transitions are one and the same 
and are referred to as simply the QCD transition. Here, we will adopt both 
these conventional assumptions and explore the consequences of the resulting 
similarity between the QCD transition and the superfluid phase transition.

Collisions of highly relativistic nuclei offer the possibility of producing
quasi-macroscopic regions of dense nucleonic matter at a sufficiently high
temperature that it might be possible to observe the QCD transition 
experimentally. In such 'heavy-ion collisions', one typically collides Pb, Au,
S or O at energies between 3 and 200 GeV per nucleon. The high energies of 
the incident nuclei mean that, in the centre of mass frame, they are highly
Lorentz contracted and resemble two pancakes approaching each other along
the beam pipe at almost the speed of light. In fact, the energies of the
incident nuclei are sufficiently high that the whole system is approximately
Lorentz boost invariant. The practical result of this is that all physical
quantities tend to be functions of proper time and hence observables are
rapidity independent.

Immediately after the collision, the two nuclei recede in opposite directions
down the beam-pipe, leaving a region of hot quark-gluon plasma between them. 
This then expands and cools through the QCD phase transition. Eventually the
energy density becomes sufficiently low that the plasma hadronises to produce
the pions, nucleons and kaons observed. Unsurprisingly by far the majority of
the products are pions since they are so much lighter than anything else.

If one is to use such heavy-ion collisions to probe the nature of the QCD
phase transition, it would clearly help to have some indication of what the
experimental consequences of QCD ought to be. Due to the difficulty of applying
conventional perturbative techniques to QCD or simulating the transition 
numerically, this is an incredibly hard problem. Nonetheless there has been 
much work on possible experimental signatures, involving observables such as
the photon and dilepton fluxes and the $ K / \pi $ ratio. There has also been 
interest in the effect of hydrodynamic instabilities during the cooling of the
plasma. Possibly the most clear signature so far considered, however, is the
deficit of neutral pions which would arise from large regions of misaligned QCD
vacuum acting as pion lasers, more usually known as dis-oriented chiral 
condensates.

Another, very natural possibility, would be to look at the baryon
density. Although the baryon number of the incident nuclei would probably
almost all 
be contained in the receeding pancakes which constitute the remnants of the
original particles, it is conceivable that there could be a significant 
baryon number density in the central region immediately after the 
transition. Since the speed of the plasma is proportional to the distance 
from the collision point, this central region is also known as the central 
rapidity region. Any baryons in it would therefore have characteristically low 
longitudinal velocities compared with the receeding nuclei.

Nonetheless, the central rapidity region is usually assumed to be baryon free,
partly on the basis of string models, although there has until now been no
work to predict the proton and neutron distributions directly \cite{werner}.
It turns out, however, that in the context of the linear sigma model the 
evolution of the baryon density has many similarities with the production
of topological defects \cite{anne}. Recent theoretical progress in this area 
\cite{whzphysrep,whznature}, supported by experiments in superfluids
\cite{lancaster,helsinki,grenoble}, means that it is now possible to 
address the question of the baryon density immediately after a heavy ion 
collision more directly. The resulting estimate of the initial baryon density
is consistent with the conventional belief that the central rapidity region 
is almost baryon free.

\section{A Model for the QCD Transition}

In order to calculate the baryon density in the central rapidity region
immediately after the transition, we must first choose a tractable model for
its dynamics. We have already assumed the chiral and deconfining transitions 
to be one and the same and second order. Let us further assume that it is 
then reasonable to use two flavour QCD instead of the full theory. If we use 
superscripts $L$ and $R$ to distinguish the left and right handed sectors of 
QCD, then the breaking of the full chiral symmetry group:-
\begin{displaymath}
U^L(N_f) \times U^R(N_f) \equiv U_V(1) \times U_A(1) \times SU^L(N_f) 
\times SU^R(N_f),
\end{displaymath}
where the vector $U_V(1)$ and axial $U_A(1)$ sub-groups correspond to 
multiplying the left and right quark spinors by equal and opposite phases,
to the residual symmetry:-
\begin{displaymath}
SU(N_f)_{L+R}
\end{displaymath}
would be described by non-vanishing expectation values for operators
of the form:-
\begin{displaymath}
{\cal M}^i_j \equiv \langle \overline{q}^i_L q_{Rj} \rangle. 
\end{displaymath}
Here $q_{Li}$ and $q_{Rj}$ are left and right handed quark spinors. 

The question of what model to use to describe the dynamics of such an 
order parameter is far from clear cut. The full QCD lagrangian
might be correct in principle but makes calculation too hard since the
interacting quanta are strongly interacting. Since below the critical 
temperature we have a good idea of what to expect phenomenologically, one 
often tries to deduce the more physically relevant weakly-interacting degrees 
of freedom and use these to construct a more tractable model. A familiar 
example occurs in condensed matter systems where one starts with strongly 
interacting atoms in a crystal lattice and transforms this into a description 
in terms of weakly interacting phonons. For QCD, the weakly interacting 
perturbative degrees of freedom are usually taken to be mesonic, or in
other words the pions. Exact transformations between the original quark-gluon 
degrees of freedom and the effective mesonic degrees of freedom are not known,
but there are two phenomenological models commonly used, the Skyrme model 
\cite{aitchison} and the $O(4)$ linear sigma model \cite{wilzcekrajagopal}.

Both of these theories roughly reproduce multi-pion scattering amplitudes to 
order $p^2$, or in other words at tree level. This is equivalent to treating
the models as classical lagrangians. If one were to treat these new 
phenomenological theories classically, one would expect the baryons to be 
solitons. The justification for a classical treatment has been much discussed 
in the literature \cite{rajagopal}. In fact, the classical approximation can
be very similar to a fully quantum mechanical, although still approximate 
treatment \cite{boyanovsky}. Physically, this is because the long wavelength 
modes of the scalar fields grow exponentially and their correlation length 
becomes larger than all other length-scales in the problem, including the 
inverse pion-masses and $1/T_C$. In other words they become more classical as 
the transition progresses. Pictorially, one can think of classical, long 
wavelength ocean swells coming to dominate over the short wavelength chop as 
the transition progresses.

Let us first consider the Skyrme model:- 
\begin{displaymath}
{\cal L}_{{\rm Skyrme}} = \frac{1}{2} \partial_\mu \phi^a \partial 
^\mu \phi_a + \frac{1}{4 \epsilon^2 \eta^2} 
\Biggl [ \,
( \partial _\mu \phi_a \partial_{\nu} \phi^{a} )^2 
- ( \partial_\mu \phi^a )^4
\, \Biggr ].
\end{displaymath}
where the vector $\phi = ( \sigma, {\bf \pi} )$ is an $O(4)$ multiplet of 
real scalar fields, the vector ${\bf \pi}$ representing the three pions and 
the $\sigma$ a sigma particle too massive to be observed at low energies.

\begin{flushleft}
\begin{figure}[tbp]
\epsfig{file=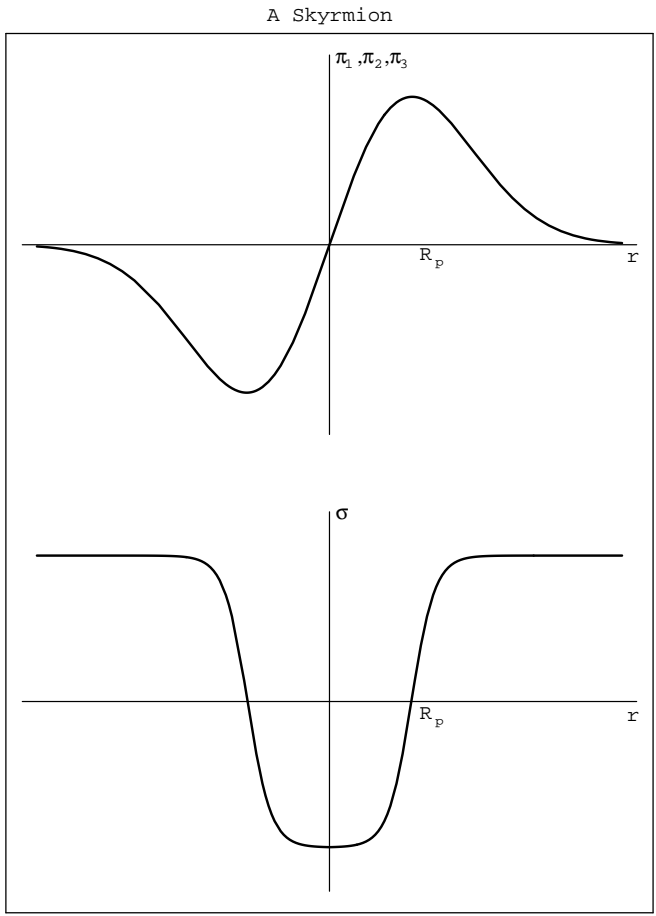, width=8.7cm}
\caption{ This configuration of the O(4) scalar field is stable for 
topological reasons. Such configurations are generically called topological 
defects and this particular example is called a Skyrmion. It corresponds to 
either a proton or neutron.}
\label{skyrmion}
\end{figure}
\end{flushleft}

Due to the extra scale coming from the four derivative term, this model 
has stable texture-like topological defects, usually called Skyrmions.
It is possible to show that these textures have baryon number one and 
spin one half and they are therefore identified with the protons and
neutrons. Unfortunately, since exact isospin invariance was assumed there is 
no distinction between the two. Since experimentally it would be far easier
to detect protons than neutrons this is potentially a serious problem.
The minimum energy solution for such a Skyrmion is of the form:-
\begin{eqnarray}
\nonumber
{\bf \pi} &=& \frac{\bf r}{r} f(r) \sin \theta(r)
\\
\sigma &=& f(r) \cos \theta(r)
\label{texture}
\end{eqnarray}
where $f$ is some function of $r$ which has to be calculated numerically by
minimising the energy to give the result shown in figure \ref{skyrmion}.

The conserved topological current associated with these textures is given by:-
\begin{displaymath}
W^\mu = - \frac{1}{12 \pi^2 \eta^4} \, \epsilon^{\mu \nu \lambda \rho}
\epsilon_{abcd} \, \phi_a \, \partial_\nu \phi_b \, \partial_\lambda \phi_c
\, \partial_\rho \phi_d,
\end{displaymath}
where $\epsilon^{\mu \nu \lambda \rho}$ is the totally antisymmetric
symbol in Minkowski space and $\epsilon_{abcd}$ is the equivalent in 
O(4) field space. The zeroth component of this gives the topological
charge density, or in other words what turns out to be the baryon number 
density:-
\begin{displaymath}
W^0 = - \frac{1}{2 \pi^2} \, \epsilon^{ijk} \epsilon_{abcd} \, 
\frac{\phi_a}{|\phi|} \, \partial_i \Biggl ( \frac{\phi_b}{|\phi|} \Biggr )
\, \partial_j \Biggl ( \frac{\phi_c}{|\phi|} \Biggr )
\, \partial_k \Biggl ( \frac{\phi_d}{|\phi|} \Biggr ).
\end{displaymath}
Since protons and neutrons are indistinguishable in this model this corresponds
to the density of nucleons minus the density of anti-nucleons. 

From the point of view of predicting the baryon density immediately after
the QCD transition, however, this model has one fatal flaw, namely the Skyrme
term itself. This term is clearly not conformally invariant since it is 
designed specifically to provide a scale for the protons and neutrons. Hence,
a model incorporating such a term can not describe a renormalisation group 
fixed point such as a phase transition and in particular couldn't describe
the chiral transition. Another way of seeing the same thing is to regard the
Skyrme term as a lagrange multiplier which fixes the vacuum expectation value
of the field. If this is constrained to be finite, clearly one can't describe 
the symmetric phase in which, by definition, the vacuum expectation value
should be zero. 

The only other alternative, without going to higher orders in some form of
derivative expansion and ending up with a model which is totally impractical
for calculation, is the $O(4)$ linear sigma model:-
\begin{displaymath}
{\cal L} = \frac{1}{2} \partial_\mu \phi^a \partial ^\mu \phi_a -
\frac{\lambda}{4} \, ( \phi_a \phi^a - v^2 ) ^2 + H \sigma.
\end{displaymath}
Again $\phi = ( \sigma, {\bf \pi} )$ is an $O(4)$ multiplet of real scalar
fields, ${\bf \pi}$ representing the pions and the $\sigma$ a sigma
particle too massive to be observed at low energies. A priori, this is a 
quantum field theory and the zero temperature values of the parameters 
$\lambda$, $H$ and $v^2$ should be chosen to give reasonable agreement 
with experiment at low energies. Given the other approximations 
associated with our choice of model, the exact experimental data used to 
choose the values of these parameters is not critical. Here the following 
values are chosen \cite{gavin,wilzcek}:-
\begin{eqnarray}
\nonumber
v &=& 87.4 {\rm MeV} 
\\
\nonumber
H &=&  ( 119 {\rm MeV} )^3
\\
\nonumber
\lambda &=& 20
\nonumber
\end{eqnarray}
which are consistent with $m_\pi =140$MeV, $m_\sigma = 600$MeV and the 
pion decay constant $f_\pi = 92.5$MeV. There are other equally valid
possibilities, however \cite{emil}. With these parameters, the sigma 
model gives a reasonable description of 
the phenomenology at less than 1GeV. The phase transition takes place at
roughly $T_C \approx f_\pi$ and low energy $\pi - \pi$ scattering amplitudes
come out about right. Probably the largest criticism of the model is that
the $\sigma$, which is far more massive than the quasi-goldstonian pions,
has never been seen. 

One advantage of the linear sigma model is that all the critical exponents of
the theory can be calculated, in both the static and dynamic renormalisation
group, in the limit of small $H$. In fact, for the three dimensional theory, 
the linear sigma model is plausibly in the same universality class as the 
$O(4)$ Heisenberg ferromagnet 
\cite{wilzcekrajagopal} whose indices have previously been calculated to 
seven loops \cite{baker}. Given the other approximations made here, this
would seem to be plenty. In the conventional notation, the critical indices
are as follows:-
\begin{eqnarray}
&\alpha& \, = \, 2 - d \nu \, = \, -0.19 \pm 0.06
\nonumber
\\
&\beta& \, = \, \frac{\nu}{2} ( d - 2 + \eta ) \, = \, 0.38 \pm 0.01
\nonumber
\\
&\gamma& \, = \, ( 2 - \eta ) \nu \, = \, 1.44 \pm 0.04
\nonumber
\\
&\delta& \, = \, \frac{ d + 2 - \eta } { d - 2 + \eta } \, = \, 4.82 \pm 0.05
\nonumber
\\
&\nu& \, = \, 0.73 \pm 0.02
\end{eqnarray}
where $d$ is the spatial dimension which we take to be three, notwithstanding
the apparent flatness induced by the approximate Lorentz boost invariance, 
since the spatial structure of the field will later turn out to be important.
If we define the relative temperature
$\epsilon = 1 - T/ T_C$ in the conventional way, then the correlation length
of the scalar field will be:-
\begin{displaymath}
\xi = \frac{ \xi_0 } { \epsilon^\nu },
\end{displaymath}
where $\xi_0 \approx 0.7$fm.
\newpage
\noindent
Similarly, the relaxation rate of the pion field may be obtained from 
dynamical renormalisation group arguments:-
\begin{displaymath}
\tau = \tau_0 \xi^z
\end{displaymath}
where
\begin{displaymath}
\tau_0 = \frac{2 \hbar} {\lambda v^2 c^2} \approx 2 \times 10^{-24}{\rm s}
\end{displaymath}
which unsurprisingly is of the order of the light crossing time of a pion.
In three dimensions, the critical index $z=d/2=3/2$.

The biggest draw-back with this model as far as we are concerned comes from
Derrick's theorem, which tells us that any renormalisable field theory
involving only scalar fields can't support stable solitons. In other words,
the fact that there is no fourth order derivative term in the sigma model
to provide a length scale for the baryons means that the configurations of 
the field which would be defects and represent the baryons are {\it not} 
topologically stable. Whereas the Skyrme model had stable objects
corresponding to protons and neutrons but could not be a good description
of the phase transition since it broke conformal invariance, the linear
sigma model is conformally invariant at the transition but does not have
stable protons and neutrons.

The most obvious solution to this dilemma would be not to require the model
to be a renormalisable field theory. Since we are looking for a 
phenomenological model rather than a fundamental field theory there is some
justification for this. In this case, however, it would not be clear exactly
what was meant by conformal invariance and it would not be easy to calculate
with whatever terms were necessary to ensure stable solitons.

Another possibility would be to exploit the fact that we are only really 
interested in a finite sized volume of quark-gluon plasma and hence have 
boundary conditions so that Derrick's theorem doesn't necessarily apply. 
Certainly, if we were to use the sigma model on a 2-sphere, for example by
exploiting the axial symmetry to reduce the problem to just the radial and
beam-pipe co-ordinates and then imposing the condition that the field should
be zero at the edge of the plasma, there would be stable solitons, even 
though these would lack a scale and the nucleon size would not be fixed.

Neither of these solutions is particularly appealing or clearly better than
the other, however. Assuming that there really is a transition and that it is 
second order, then ideally one would like a renormalisable theory which is 
conformally invariant at the critical point and conserves baryon number 
written in terms of the pion fields. It is not clear, however, that this is 
possible since the bosonic theory does not include either glueballs or all the 
quark flavours. It could also be the case that in reality there is a smooth 
cross-over rather than a real phase transition, in which case the theory 
wouldn't have to be conformally invariant at the critical point.

Here, however, we have assumed a second order transition in two flavour 
QCD. In order to describe the transition we therefore have to use the sigma
model. In both the Skyrme and sigma model, however, the $O(4)$ scalar field
represents the same physical degrees of freedom \footnote{Although in the case
of the Skyrme model there is some ambiguity in the choice of which of the 
components is the massive sigma.}. We know from the Skyrme model what 
configuration of the pion and sigma fields corresponds to a baryon and hence
the same configuration ought also to correspond to a baryon in the sigma model.
Indeed, if we consider the sigma model in thermal field theory, at low 
temperatures, the field would effectively be confined to the vacuum manifold,
the texture configurations would be effectively stable and the model would be 
equivalent to the Skyrme model. We will therefore assume that the QCD phase
transition can be described by the linear sigma model and that the protons
and neutrons are represented by Skyrmion-like configurations of the field 
\ref{texture} notwithstanding the fact that they are not topologically 
stable.

In fact, Skyrmion-like configurations of the scalar field will tend to 
collapse at the speed of light. If we were to take this seriously this would
imply violation of baryon number. This is clearly a flaw in the sigma model
and presumably arises since this model doesn't contain all of the relevant
physics. Is it, however, serious for the dynamics of the phase transition?
Certainly, the presence of topological defects can produce non-perturbative
effects which can, for example, change the critical temperature \cite{zumbach}.
The time-scales for this process, however, are such that the rate of texture
decay, or in other words baryon violation due to the inadequacy of the model,
is always slower than the time-scale for breaking the symmetry.
The time-scale for the symmetry to be broken is of order $2 \hbar / \lambda 
v^2 c^2 \approx \xi_0 / c$. This should be compared with the minimum 
texture collapse time of $\xi / c$. Since a texture corresponding to a proton
or a neutron will always be larger than the cold coherence length, it is
safe to assume that textures will take longer to decay than the time available
during the course of the transition. In fact, detailed studies of texture 
dynamics suggest that texture unwindings may be quite rare and even less 
likely than a simple time-scale argument suggests \cite{borrill}. It will 
therefore be assumed that the of the O(4) field is unlikely to be much 
influenced by the texture unwinding events which would describe proton / 
neutron decay. 

In conclusion, in order to calculate the baryon density immediately after
a heavy ion collision, we will use the non-linear sigma model with critical
exponents calculated in the $H=0$ limit to model the dynamics and assume
that baryons correspond to configurations of the $O(4)$ scalar field 
which look like Skyrmions.

\section{Initial Conditions and Hydrodynamics}

Let us now consider the initial conditions.
If the energy released in a heavy-ion collision is roughly equivalent to 
that produced in nucleon-nucleon collisions then one expects the energy
density immediately after the collision to be of order 3 GeV / fm$^3$.
Unlike many cosmological phase transitions, however, QCD is a strongly coupled
system and, in the sigma model, the dimensionless coupling constant $\lambda$ 
must be of order 20 in order to approximate low energy pion cross-sections. 
At these sort of energy densities, this means that the interaction time-scale 
is likely to be far shorter than the cooling rate and the deconfined 
quark-gluon plasma is likely to come rapidly into thermal equilibrium at a 
temperature of a few hundred MeV. 

For example, if following Bjorken \cite{bjorken} one assumes the initial 
energy density to be somewhere between 1 and 10 GeV / fm$^3$ and distributes
this energy at about 400 MeV per quantum, then the mean free path works
out to be roughly:-
\begin{displaymath}
\lambda_{mfp} \approx \biggl ( \frac{10 {\rm mb} } { \sigma_{int} } \biggr )
\times (0.05 - 0.5) {\rm fm}
\end{displaymath}
with a corresponding thermal equilibration time of the order of 1 fm / c. 
Thus, while the usual assumption of an initial thermal state is likely to be 
wrong in the comparatively weakly coupled cosmological models, it may be a 
reasonable approximation for a heavy-ion collision. 

Typically people assume you reach 200 and 300 MeV at between 1 and 4 fm / c 
\cite{emil}. Here, as an initial condition, we will assume that the 
quark-gluon plasma is in thermal equilibrium at a temperature of 200-300 MeV 
at a proper time $\tau_I \approx 1$ fm / c after the collision. 

There is in fact another reason not to try to use the sigma model 
at times less than 1 fm / c. If we were to treat the sigma model as a 
phenomenological quantum theory rather than as a classical field theory, 
some momentum cut-off, $\Lambda$ would be needed. 
Clearly $2 m_\pi < m_\sigma < \Lambda$ in order to allow fluctuations of
the field on the scale of the pion compton wavelength. What is less obvious
is that, with $\lambda$ of the order of ten or twenty, a cut-off much
larger than 1 GeV leads to a negative effective coupling, thus constraining
$600 {\rm MeV} < \Lambda < 1 {\rm GeV}$. This is equivalent to a length-
scale cut-off of the order of 0.2fm and hence the model wouldn't make much
sense on time-scales much less than 1 fm / c. It is safe to assume that we 
should also not take the classical theory seriously on such length scales.

At these sorts of energies, in the centre of mass frame, both the incoming
nuclei appear highly Lorentz contracted into pancake shapes. In addition,
experiments see uniform particle production as a function of rapidity, at 
least from the collision or central rapidity region. Both of these facts imply
an approximate Lorentz boost invariance and consequently that physical 
quantities should depend only on the proper time $\tau = \sqrt{t^2-x^2}$, 
where $x$ is the co-ordinate along the beam-pipe with zero at the collision 
point. The consequence of this is that initially at least the expansion of the 
plasma will be linear along the beam pipe. This should be true for times ( or 
distances from the collision axis ) of the order of a nuclear 
radius, or $t \ll 1.2(A_1+A_2)^{1/3} \approx 7$ fm / c for lead or uranium,
where $A_1$ and $A_2$ are the mass numbers of the colliding nuclei. At later 
times, one would expect three dimensional rather than linear expansion but 
since this volume is at least as big as the region in which protons and 
neutrons are likely to be formed we ignore this.

Since we are assuming, as is conventional, that the sigma model may be treated
classically in the context of the chiral transition, it follows that during
the initial linear expansion of the plasma, the temperature falls off like:-
\begin{displaymath}
T = T_I \Biggl ( \frac{\tau_I}{\tau} \Biggr )^{\alpha},
\end{displaymath}
where $\alpha$ depends on the speed of sound in the plasma and is 1/3 
in the case of an ultra-relativistic plasma \cite{bjorken}.
The fact that the temperature can only depend on $\tau$ implies that the
plasma is hottest just behind the receeding pancakes and coolest in the 
central region, and cools from the inside outwards, somewhat like a 
baked-alaska.

\begin{flushleft}
\begin{figure}[htbp]
\epsfig{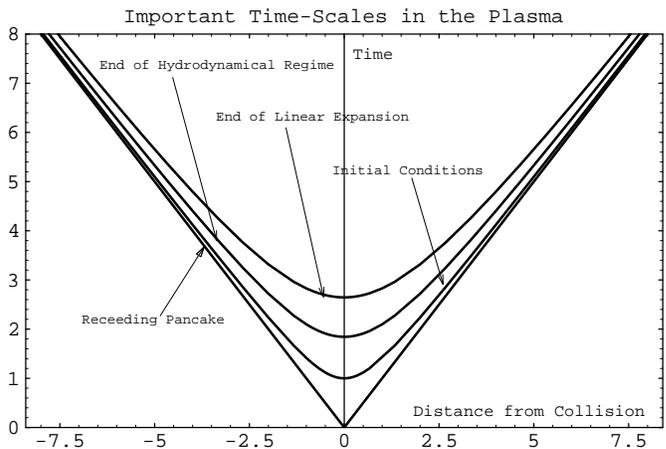}
\caption{A pictorial representation of the evolution of the plasma in space 
time. The remains of the incident heavy ions receed down the beam-pipe on 
light-like trajectories and all physical quantities and events are either 
specified or take place on space-like hypersurfaces of constant proper time.}
\label{echelles}
\end{figure}
\end{flushleft}

Clearly, the plasma and the sigma model can only be treated classically while
there is more than one pion per pion compton wavelength:-
\begin{displaymath}
kT > kT_I \Biggl ( \frac{\tau_I}{\tau_H} \Biggr ) ^\alpha 
= \frac{ m_\pi c^2 }{\lambda_\pi^3 },
\end{displaymath}
or in other words up until the proper time $\tau_H$ such that:-
\begin{displaymath}
\tau_H \le \tau_I \Biggl [ 
\frac{ k T_I \lambda_\pi^3 } { m_\pi c^2 }
\Biggr ]^{ 1 / \alpha }
\end{displaymath}
For an initial temperature of about 200 MeV / fm$^3$ this works out to be 
about $3.4 \tau_I$ if one takes the pion Compton wavelength to be one fermi and
proportionally larger if one takes the value 1.5 fm. This is worth noting, 
since with the smaller value it is conceivable that the plasma would hadronize
sufficiently early to be of dynamical interest. The relation between these 
time-scales is shown in figure \ref{echelles}.

\section{Zurek Scenario in a QCD Plasma}

To summarise, our picture of a heavy ion collision in the laboratory frame is 
of two highly lorentz contracted nuclei colliding to produce an approximately
boost invariant plasma, initially in local thermal equilibrium. This then
cools from the inside out, expanding initially linearly, with correlation 
domains which grow as the phase transition is approached. This we will describe
using a classical treatment of the linear sigma model, with critical exponents
calculated in the $H=0$ limit. In order to predict the number of protons and 
neutrons produced in the central rapidity region, we need to know how many 
Skyrmion like configurations of the pion field will be produced in traversing 
the phase transition.

Although they will not actually be topologically stable in the sigma model,
and will only become topologically stable objects when the plasma has cooled
sufficiently below the transition that the field is almost always on the 
vacuum manifold and the Skyrme model is appropriate, the formation of these
Skyrmion like field configurations will presumably be very similar to the 
formation of regular topological defects by the Kibble mechanism \cite{tom}.
In fact, counting the number of Skyrmions produced in this symmetry-breaking 
phase transition is equivalent to counting the number of topological defects 
produced in many other phase transitions, including those in superfluid 
helium which have been used to experimentally test our ideas concerning
defect formation. {\it This is the crucial point which we exploit here in
order to estimate the baryon density in the central rapidity region.}

In the present context, the Kibble mechanism would work as follows.
During the transition, the O(4) scalar pion field begins to fall from the 
false ground-state into the true ground-state, choosing a point on the 
ground-state manifold at each point in space. We will assume that the
bias induced by the sigma term is small and that the point on the vacuum
manifold is chosen approximately at random. Our sigma model has a second
order phase transition so this collapse to the true ground-state will occur
by phase separation, the resulting field configuration being one of
domains within each of which the scalar field has relaxed to a constant
ground-state value.

\begin{flushleft}
\begin{figure}[tbp]
\epsfig{file=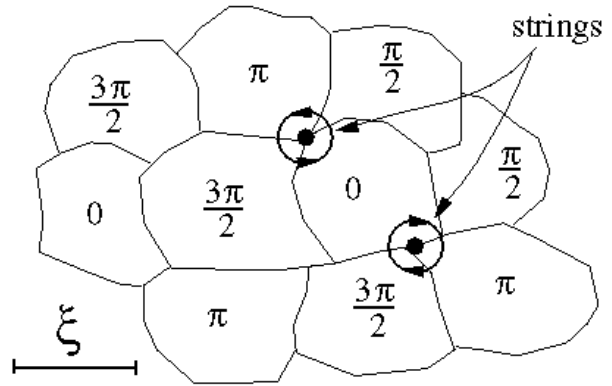, width=8.7cm}
\caption{In the conventional Kibble mechanism, topological defects, most
often vortices of some kind, are formed between the domains of correlated
field.}
\label{domains}
\end{figure}
\end{flushleft}

In the conventional Kibble mechanism, one then argues that continuity and 
single valuedness will sometimes force the field to remain in the false 
ground-state between some of the domains. This requires at least one
zero of the field which would have topological stability and
characterise a defect. The density of defects is then closely linked to the
number of domains as shown in figure \ref{domains}.

In the case of the linear sigma model, however, it would be perfectly
possible for the field to have Skyrmion like winding but still not be a 
defect since the field could unwind inside the surface over which the winding
is computed. Counting protons and neutrons with the conventional topological
current then might be hard. We can however still exploit heuristic evidence
from simulations which indicates that there will be one real Skyrmion 
configuration formed for every twenty-five to a hundred domain sizes
\cite{borrill}. Calculating the proton-neutron density then becomes
a case of finding how many correlation domains are formed within the plasma.

The question of how big the correlation domains are when defects are formed
is the subject of the Zurek scenario for the formation of topological defects
\cite{whznature}. This scenario provides an estimate of the initial defect 
density immediately after the phase transition in a particular case of the
Kibble mechanism, namely a rapid quench through a second order phase 
transition. Although this scenario arose through considering the 
possible formation of cosmic strings in the early universe, it has since been 
tested using the superfluid transitions in both $^4$He \cite{lancaster} and 
$^3$He \cite{helsinki,grenoble} and has so far been consistent with all 
observations. In fact the systems in which the scenario has been tested are 
all strongly coupled in some sense and are hence far closer to the QCD 
transition in terms of their dynamics than they are to the sorts of 
cosmological phase transitions typically considered.

\begin{flushleft}
\begin{figure}[tbp]
\epsfig{file=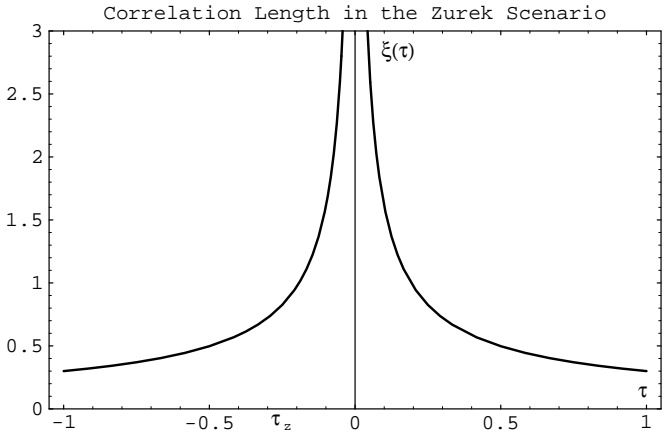, width=8.7cm}
\caption{Pictorial representation of the evolution of the correlation length
according to the Zurek scenario. Initially the plasma is sufficiently close
to equilibrium that the correlation length is able to track the quench quite
well and the correlation length stays quite close to the equilibrium value.
Eventually, as the transition is approached, however, it must go out of 
equilibrium at a particular proper time, $\tau_Z$.}
\label{correl}
\end{figure}
\end{flushleft}

In the Zurek scenario, the prediction of the defect density formed during 
a symmetry breaking phase transitiondepends on the phenomenon of critical 
slowing down. As the plasma cools and approaches the phase transition, the 
correlation length grows and the relaxation rate of the pion field gradually 
decreases like $\tau=\tau_0 \xi^z$. At some stage during the cooling process 
therefore the relaxation time-scale of the pion field will become longer than 
the time-scale on which the plasma is cooling. In other words, the pion field 
will no longer be in thermal equilibrium and will be unable to keep up with 
the cooling of the plasma \footnote{Actually, the relaxation time of the field
is mode-dependent so that, at least in a Gaussian approximation, different 
modes of the field would go out of equilibrium at different times. Remarkably,
however, this turns out not to affect alter a naive estimation of the defect 
density \cite{glyknray}.}. This is shown schematically in figure \ref{correl}.

Hence, during the cooling of the QCD plasma, two regimes can be distinguished.
Initially, sufficiently far away from the critical temperature the relaxation
time-scale is much smaller than the time on which the cooling is proceeding
and the pion field can maintain itself in local equilibrium with a correlation
length $\xi(T)$. By contrast, when the plasma has cooled to a relative
temperature in the vicinity of the phase transition, the pion field is 
effectively frozen compared to the time-scale on which the plasma is cooling. 
Thus, whatever the configuration and correlation length of the field at this
relative temperature, it will be frozen in until after the phase transition 
when the field is again in thermal equilibrium. The time at which the plasma 
moves from one regime to another we will refer to as the Zurek time 
$\tilde{t}$. Quantities evaluated at this time will be denoted by a tilde.

Of course, by the time the field comes back into equilibrium after the phase
transition, the it may not be appropriate to treat the pion field classically
\footnote{Indeed, it will turn out that the pion field goes out of equilibrium
only marginally before the end of the classical region.}. The important point,
however, is that so long as the field behaves classically until the Zurek time,
when the field first goes out of equilibrium, the length scale which is 
imprinted on the plasma is the coherence length at the Zurek time.

As an aside, we note that even if the chiral and deconfining transitions
are actually separate and yet occur fairly close together, or even if there 
is not real transition but merely a slow change of phase, so long as there is 
a critical slowing down of the field relaxation time only one length will be 
frozen in. It may be possible to study the question of two transitions further 
in $^3$He or two dimensional networks of superconducting wire.

Unlike cosmological phase transitions, however, which are always assumed to
occur homogeneously, the quark gluon plasma cools from the inside out somewhat
like a baked-alaska. In other words the second order transition into the 
symmetry broken phase occurs by the propagation of a temperature front. The
most striking example of this occurs in nematic liquid crystals where the
phase front moves particularly slowly, but the transition proceeds in the same
way for many different systems. The speed of this front may be calculated as 
follows.

Since the temperature depends on the proper time $\tau = \sqrt{c^2t^2-x^2}$
as $T=T_I ( \tau_I / \tau )^\alpha$, at any given time, the temperature 
gradient will be:-
\begin{displaymath}
\frac{\partial T}{\partial x} = T_I \alpha \tau_I^\alpha (c^2 t^2 - x^2 )
^{ - \frac{\alpha + 2}{2} } x
\end{displaymath}
Similarly, for any given position, the rate of change of temperature with 
respect to conventional time will be:-
\begin{displaymath}
\frac {\partial T}{\partial t} = - c^2 \alpha T_I \tau_I^\alpha (c^2 t^2 - x^2
) ^{ - \frac{\alpha + 2}{2} } t
\end{displaymath}
In other words, the length and time scales associated with temperature
change will be:-
\begin{eqnarray}
\frac{1}{\lambda_T} &=& \frac{1}{T} \frac {\partial T}{\partial x} =
\frac{ \alpha x } {\tau^2}
\nonumber
\\
\frac{1}{\tau_Q} &=& \frac{1}{T} \frac {\partial T}{\partial t} =
\frac{ c^2 \alpha t}{\tau^2}
\\
\nonumber
\end{eqnarray}
Hence, in the centre of mass frame, this front will propagate with speed:-
\begin{displaymath}
v_T = \frac{\lambda_T}{t_Q} = \frac{c^2 t}{x}
\end{displaymath}
However, for the region inside the plasma, $ x < ct $ except very close to 
the receeding crepes where the exact boost invariance breaks down. Hence
the acausal temperature front will always move at least as fast as the 
speed of light. In other words, the phase front propagates not only faster
than the fluid which moves with a bulk velocity along the beam pipe of $x/t$
but also faster than the sound speed which is relevant for the equilibration
of the plasma. Effectively this means that there is no difference between
this baked-alaska cooling and the homogeneous case as far as the formation
of topological defects is concerned \cite{tomgrisha}.

Let us therefore compute the relative temperature at which the textures
will be frozen in for some point within the plasma with co-ordinates $(x,t)$
in the centre of mass frame. Clearly, the field will go out of equilibrium 
when the time-scale associated with the cooling, $t_Q$, is equal to the 
relaxation rate of the plasma, $t_R$. As earlier, 
\begin{displaymath}
\frac{1}{\tau_Q} = \frac{ c^2 \alpha t}{\tau^2}
\end{displaymath}
and 
\begin{displaymath}
t_R = \frac{t_0}{\epsilon^{\nu z}} 
\end{displaymath}

Suppose we consider a point in the plasma a fixed fraction $\sqrt{1-f^2}$ of 
the distance from the point of impact to the position of the crepes at the 
outermost extremity of the plasma with $x \approx ct$. At any time $t$, this 
point $(t,\sqrt{1-f^2}ct)$ will have the proper time
\begin{displaymath}
\tau = \sqrt{c^2 t^2 - x^2} = \sqrt{ c^2 t^2 - c^2 t^2 (1-f^2) } = ctf
\end{displaymath}
and consequently the quench time-scale at this point will be 
$t_Q = ft / \alpha$. Since we know how to calculate the relative temperature
$\epsilon$ as a function of proper time $\tau$ we can relate $ft$ to 
the relative temperature at the point $(t,\sqrt{1-f^2}ct)$ in the plasma
as:-
\begin{displaymath}
tf = \frac{ \tau_I}{c} \Biggl [ 
( 1 + \epsilon ) \frac{T_C}{T_I}
\Biggr ]^{1 / \alpha}
\end{displaymath}

Equating the two time-scales then yields:-
\begin{displaymath}
\frac{\alpha c t_0}{\tau_I} \Biggl ( 
\frac{T_I}{T_C}
\Biggr )^{ 1 / \alpha } 
\approx \, \tilde{\epsilon}^{\nu z} \Biggl ( 
1 + \frac{\tilde{\epsilon}}{\alpha}
\Biggr )
\end{displaymath}
This can be solved numerically for the relative temperature $\tilde{\epsilon}$
which is frozen into the O(4) field when it goes out of equilibrium and the
length-scale imprinted on the plasma will be:-
\begin{displaymath}
\tilde{\xi} = \frac{ \xi_0 } { \tilde{\epsilon}^\nu } 
\end{displaymath}
It will be seen that this relative temperature does not depend on $f$, the 
position in the plasma. In other words the plasma goes out of equilibrium
at the same relative temperature everywhere and a single length-scale is 
imprinted on the plasma. 

The approximate Lorentz boost invariance of the plasma implied that physical
quantities such as the initial conditions were specified on space-like 
hypersurfaces of constant proper time, as shown in figure \ref{echelles}.
Although the plasma goes out of equilibrium at different times depending on 
how close to the receeding crepes it is, the freezing in of topological 
defects actually occurs on a space-like hypersurface of constant proper time.

Since for $x=0$ the proper time is equal to $ct$, using the fact that
$T=T_I ( \tau_I / \tau )^\alpha$ one finds that $\tau_C = ( T_I / T_C )^{ 1 / 
\alpha } \tau_I$. Substituting our expression for temperature in terms of
time at $x=0$ into our definition of relative temperature $\epsilon = T / T_C
- 1$ one finds:-
\begin{displaymath}
\tau = \tau_I \Biggl [ 
\frac{T_I}{T_C} \frac { 1 } { 1 + \epsilon }
\Biggr ]^{ 1 / \alpha}
\end{displaymath}

Rather than solving for $\tilde{\epsilon}$ numerically, however, it is 
possible to obtain an approximate solution by specialising to the case
of $x=0$ or equivalently $f=1$ and using a slightly different criterion
for when the pion field goes out of equilibrium, namely that the relaxation
time is equal to the time remaining until the phase transition occurs:-
\begin{displaymath}
t_C - \tilde{t} = t_R
\end{displaymath}

The criterion for going out of equilibrium then becomes:-
\begin{displaymath}
t_I \Biggl( \frac{T_I}{T_C} \Biggr ) ^{ 1 / \alpha } 
\Biggl [
1 - \Biggl ( \frac {1} { 1 + \tilde{\epsilon} } \Biggr ) ^{ 1 / \alpha }
\Biggr ]
= \frac{t_0}{ \tilde{\epsilon}^{\nu z} }
\end{displaymath}
Exploiting the fact that $\epsilon$ is likely to be small near the phase
transition when the field actually goes out of equilibrium and in any
case will certainly be smaller than the inital value of roughly $0.25$ we
can series expand $( 1 + \tilde{\epsilon} ) ^{- 1 / \alpha}$ to give:-
\begin{displaymath}
\tilde{\epsilon} \vert_{x=0} = \Biggl [
\alpha \frac{t_0}{t_I} \Biggl ( \frac{T_C}{T_I} \Biggr ) ^{ 1 / \alpha }
\Biggr ]^{\frac{1}{1+ \nu z}}
\end{displaymath}
Since we know that the plasma freezes out at the same relative temperature
everywhere, and in particular it freezes out on a specific space-like 
hypersurface of constant proper time, we may as well use this approximate 
solution rather than solving the previous equation numerically.

Using this approximate solution for the relative temperature $\tilde{\epsilon}$,
we can calculate the length scale which is frozen into the plasma during the
phase transition, the Zurek length. Since we know that volume of the QCD plasma
at any particular time we can then calculate how many correlation volumes are
frozen into the plasma according to the Zurek scenario and hence how many
protons and neutrons we would expect to see in the central rapidity region.

Since the initial conditions are not well known the results for a variety of
plausible initial temperatures and proper times are shown:-
\vspace{0.2cm}\\
\begin{tabular}[h]{|l||r|r|r|r|r|}  
\hline \hline
$\tau_I / fm$ & 1 & 1 & 2 & 4 & 4 \\
$T_I$ / MeV/fm$^3$ & 200 & 300 & 250 & 200 & 300 \\
$c t_c$ / fm & 2.0 & 6.6 & 7.6 & 7.8 & 26.4 \\
$\epsilon_I$ & 0.25 & 0.88 & 0.56 & 0.25 & 0.88 \\
\hline \hline
$\tilde{\epsilon}$ & 0.36 & 0.20 & 0.19 & 0.19 & 0.10 \\
$\tilde{\xi}$ / fm & 1.5 & 2.3 & 2.4 & 2.4 & 3.8 \\
Plasma Volume / fm$^3$ & 246 & 1170 & 1386 & 1417 & 6098 \\
Number of Domains & 18 & 24 & 25 & 26 & 28 \\
\hline \hline
\end{tabular}
\vspace{0.2cm}\\
where we have taken the QCD plasma to be a cyliner of width 14fm and length equal to 
$2 c \tilde{t}$. Although both the Zurek proper time, or equivalently the frozen 
in domain size, and the volume of the plasma at this time are quite sensitive to 
the initial conditions, which are not well known, the total number of correlation 
volumes frozen into the plasma is much less sensitive since it is the ratio 
$\tilde{V} / \pi \tilde{\xi}^3$.

In other words there will be somewhere between 15 and 30 coherence volumes frozen 
into the plasma. From simulations,however, we expect to get roughly one 
skyrmion configuration per every 25 - 100 domains. In order to create a 
proton-antiproton pair then \footnote{Creation of a single proton would 
violate baryon number conservation.} we would need to have between 50 and 200 
coherence volumes and also be lucky enough that the pair didn't immediately 
annihilate. Avoiding annihilation is not entirely implausible since the 
distance between the nucleon and anti-nucleon must be at least as great as the
Zurek length which in this case is slightly greater than the proton Compton 
wavelength. Forming a sufficiently large number of domains that the resulting proton
and neutron density would be observable on the other hand is far more difficult.
Roughly speaking, however, this result indicates that the central 
rapidity region of the plasma is likely to remain free of protons and neutrons
as is usually thought. 

With the smallest possible length-scale imprinted, namely the pion compton 
wavelength, the situation is somewhat better and one could conceivably see
a reasonable number of protons and neutrons. However, this corresponds to 
almost no domain growth and in this context DCCs would also be ruled out.

It is possible, however, that we might be extraordinarily lucky since the
predicted values are necessarily rather approximate and do not categorically
exclude the possibility of a detectable baryon number in the central rapidity
region. Also, the freeze out occurs at a sufficiently large proper time that 
it is debatable whether we are still justified in treating the model 
classically. The model is already known to be flawed since it does not 
conserve baryon number and the results are somewhat suspect because of this. 
It is certainly possible to do a slightly better calculation to include the 
quantum mechanical aspects of the theory, but this would not cure the problem 
of baryon number conservation. It seems likely that the only way to make a 
significant improvement would be to improve the model somehow, and this would
probably make further analytic work intractable and necessitate a simulation.

\section{Conclusions}

The conventional wisdom is that the central rapidity region of a heavy-ion
collision in which a quark-gluon plasma is produced should be almost baryon
free. A direct estimate of this initial baryon density in the context of
the $O(4)$ linear sigma model shows this belief to be well founded and 
entirely consistent with the production of dis-oriented chiral condensates.

However, a non-zero baryon density is not ruled out by many orders of 
magnitude and the result is somewhat sensitive to the initial conditions. It 
is therefore conceivable that with a slightly larger volume of QCD plasma 
one might occasionally produce a small but non-zero baryon density. To 
determine whether it is worthwhile looking for this experimentally further 
theoretical work is necessary.

\section*{Acknowledgements}

The author would like to thank the Swiss National Science Foundation and the 
DAMTP of the University of Cambridge for financial support during this work. 
I would like to thank J-P. Blaizot for the original idea, N.S. Manton and 
N. Turok for several useful conversations and G. Karra, M. Martin and R.J.
Rivers for their knowledge of the QCD literature. I would also like to thank
L.M. Bettencourt for comments on the manuscript.

\end{document}